\documentclass[aps,pre,amsmath,twocolumn,superscriptaddress,showpacs]{revtex4-2}
\usepackage{calc}
\usepackage{amsmath}
\usepackage{amsfonts}
\usepackage{amssymb}
\usepackage{graphicx}
\usepackage{hyperref}
\usepackage{soul}
\usepackage{mathtools}
\usepackage{verbatim}
\usepackage{epstopdf}
\usepackage{subfigure}
\usepackage{latexsym}
\usepackage{dcolumn}
\usepackage{epsf}
\usepackage{float}
\usepackage[table]{xcolor}
\usepackage{multirow}
\usepackage[toc]{appendix}
\usepackage{color} 
\usepackage{graphicx}% Include figure files
\usepackage{graphicx,epstopdf}
\usepackage{xcolor}
\usepackage{dcolumn}% Align table columns on decimal point
\usepackage{bm}% bold math
\usepackage{mathrsfs} % script-like, curvy letters.
\usepackage{mathrsfs}
\begin{document}
\title{How Mobile Oscillators Reshape Dynamical Robustness in Complex Networks}
%Aging Transition among Mobile Oscillators: How Mobility Reshapes Dynamical Robustness}

\author{Sayantan Nag Chowdhury}
\email{jcjeetchowdhury1@gmail.com}
\affiliation{Department of Mathematics, Nalanda University, Rajgir, Bihar 803116, India}

\author{Srilena Kundu}
\affiliation{Helmholtz Institute for Functional Marine Biodiversity at the University of Oldenburg (HIFMB), Oldenburg, Germany}
\affiliation{Alfred-Wegener Institute, Helmholtz Center for Polar and Marine Research, Bremerhaven, Germany}

%\subject{epidemic spreading, complex network}

%\keywords{Complex network, epidemic spreading, final outbreak size, test-kit}
%\thanks{These two authors contributed equally}
	
%\thanks{Corresponding Auhtor: Chittaranjan Hens}
%\email{chittaranjanhens@gmail.com}

%%%% Abstract text to be placed here %%%%%%%%%%%%
\date{\today}

\begin{abstract}
While the robustness of collective dynamics in static networks is well established, how agent mobility and network topology jointly shape dynamical robustness remains largely unexplored. In this letter, we investigate the robustness of mobile Stuart-Landau oscillators performing random walks on networks and interacting diffusively when co-located. In the rapid-mobility regime, we reduce the dynamics to a mean-field framework linking network degree statistics to an effective global coupling strength. We show analytically that increased degree heterogeneity in the network enhances the effective coupling strength, and strictly lowers the critical threshold of functional degradation for amplitude death. Numerical simulations confirm the predicted heterogeneity-dependent ordering of the transition thresholds, showing an earlier collective collapse in scale-free networks than in random or regular networks. Our results establish mobility and network heterogeneity as key determinants of robustness in mobile oscillator populations, with implications for resilient biological and engineered collectives.
\end{abstract}

\maketitle

\maketitle

%\section{Introduction}

\par Collective oscillations are a fundamental mechanism through which interacting units, from rhythmic cellular populations to engineered robotic swarms, coordinate their macroscopic behavior. A central question in nonlinear dynamics is {\it how much functional deterioration a system can tolerate before its collective rhythm collapses}. This question is captured by the \emph{aging transition} \cite{daido2004aging,majhi2024dynamical}, in which an increasing fraction of components loses the intrinsic ability to sustain oscillations, eventually driving the population from a dynamically active state to collective quiescence. Previous studies have extensively investigated this transition in oscillator populations with static interaction networks \cite{tanaka2014dynamical,daido2011dynamics,kundu2018resumption, thakur2014time,kundu2017survivability, ratas2016macroscopic,sun2017aging,kundu2018augmentation, tanaka2012dynamical,biswas2025effect,kundu2019chemical,ray2019aging,gambuzza2017analysis,morino2011robustness, kundu2021persistence, liu2016enhancing}, where each oscillator remains associated to a fixed node in the network and interacts with a persistent set of neighbors.

\par In many natural and artificial systems, however, interactions arise not from permanent links but from motion \cite{holme2012temporal,frasca2008synchronization,chowdhury2020distance, ghosh2022synchronized, holme2004structure, frasca2012spatial, Sar_2022, fujiwara2016synchronization, dimitriadis2010tracking, chowdhury2019extreme, sudhashu2019, PhysRevLett.96.088702, nag2020cooperation, levis2017synchronization, meloni2009effects}. Motile cells encounter different cellular environments as they migrate; autonomous robots repeatedly enter and leave communication zones; and mobile sensing platforms encounter different peers as they navigate constrained environments. In such systems, the interaction network of each unit changes continuously over time. This dynamic reshuffling raises a fundamental question: {\it how does mobility alter the robustness of collective oscillations when an increasing fraction of the population becomes functionally inactive (unhealthy)?}

\par Although dynamical robustness in a system of progressive functional degradation has been extensively studied in networks with static interactions, the role of mobility in shaping the transition remains largely unexplored. To address this, in this letter we use a simplified model of mobile Stuart-Landau oscillators \cite{kuramoto2003chemical} that perform random walks on networks and interact diffusively with oscillators occupying the same node. This framework brings together three ingredients that are usually considered separately: {\it intrinsic nonlinear oscillatory dynamics, mobility-induced temporal interactions, and progressive functional degradation}. By comparing regular, random, and heterogeneous network topologies, we ask how the organization of the mobility environment influences the collective response to deterioration.

\par We find that mobility makes network heterogeneity a key determinant of the system's robustness. Rather than maintaining a fixed neighborhood, each oscillator continuously samples different parts of the underlying network, causing structural heterogeneity to emerge as an effective modulation of the collective interaction strength. Remarkably, stronger mobility-mediated coupling does not necessarily enhance dynamical robustness. Because coupling transmits the dynamical state of both active and degraded agents, increased effective coupling can also facilitate the influence of quiescent (unhealthy) agents on the active (healthy) population. As a result, heterogeneous mobility environments can suppress collective oscillations at a smaller fraction of degraded agents than more homogeneous environments. Our results demonstrate that the dynamical robustness of mobile oscillator populations is governed not by network topology alone, but by the interplay between mobility, network heterogeneity, and functional degradation.

\begin{figure*}[htp]
\centerline{\includegraphics[width=1.0\textwidth]{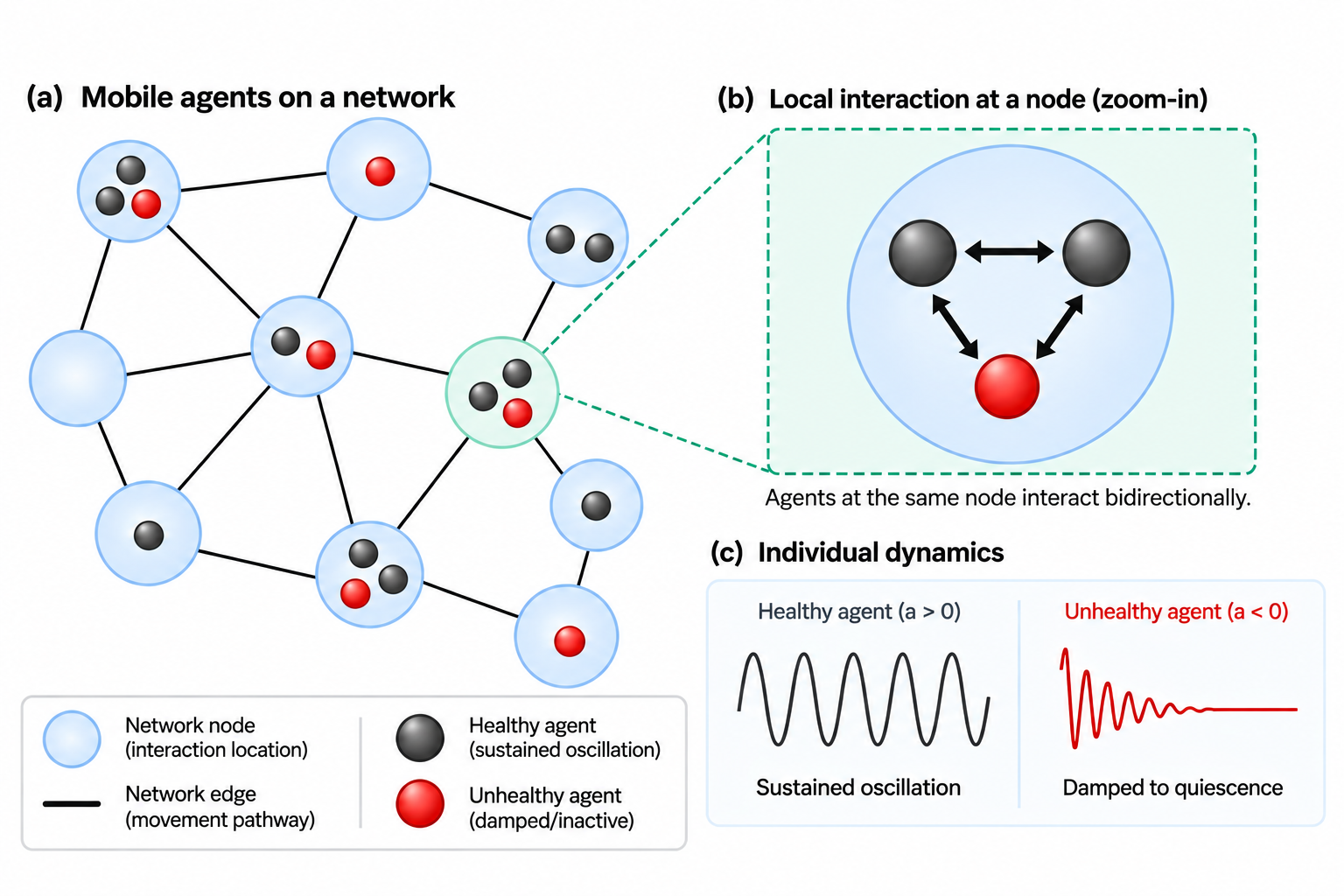}}
\caption{\textbf{Dynamically active (healthy) and inactive (unhealthy) mobile agents on a network.}
(a) Mobile agents perform random walks on an underlying network, with each node representing an interaction location. Healthy and unhealthy agents are shown in dark gray and red, respectively. 
(b) Zoomed view of a representative node containing two healthy and one unhealthy agent. Co-located agents interact bidirectionally through diffusive coupling. 
(c) Individual dynamics of healthy ($a>0$) and unhealthy ($a<0$) agents, showing sustained oscillations and decay to quiescence, respectively.}
\label{Picture1}
\end{figure*}

% ============================================================
% Analytical critical fraction for the mobile-agent model
% ============================================================

%\section{Analytical Critical Fraction of Aged Agents}

%\subsection{Microscopic model}

We consider a population of $N$ mobile Stuart-Landau oscillators distributed over a network consisting of $n$ nodes. The complex state of agent $i$ is denoted by $Z_i(t)$, and its position on the network is denoted by $x_i(t)$.  Each agent is either healthy ($a_i>0$) or unhealthy ($a_i<0$).

The dynamics of an individual agent are given by
\begin{equation}
\frac{dZ_i}{dt}
=
\left(a_i+i\omega-|Z_i|^2\right)Z_i
+
k
\sum_{j\neq i}
\mathbf{1}_{\{x_i=x_j\}}
\left(Z_j-Z_i\right),
\label{eq:microscopic_model}
\end{equation}
where $k \geq 0$ is the microscopic coupling strength, $\omega$ is the natural angular frequency, and $\mathbf{1}_{\{x_i=x_j\}}$ is an indicator function that equals one when agents $i$ and $j$ occupy the same network node and zero otherwise (See the schematic representation in Fig.~\eqref{Picture1}). For an isolated oscillator ($k=0$), the amplitude equation follows from Eq.~\eqref{eq:microscopic_model}:
\begin{equation*}
\frac{d|Z_i|}{dt}
=
\left(a_i-|Z_i|^2\right)|Z_i|.
\end{equation*}
Consequently, an isolated healthy oscillator with $a_{\mathrm{healthy}}>0$ has a stable oscillation amplitude
\begin{equation*}
|Z_i|=\sqrt{a_{\mathrm{healthy}}},
\end{equation*}
whereas an isolated {\color{black}unhealthy} oscillator with $a_{\mathrm{unhealthy}}<0$ approaches the quiescent state $Z_i=0$.

% ------------------------------------------------------------
%\subsection{Effect of rapid mobility}

\begin{figure}[htp]
\centerline{\includegraphics[width=0.5\textwidth]{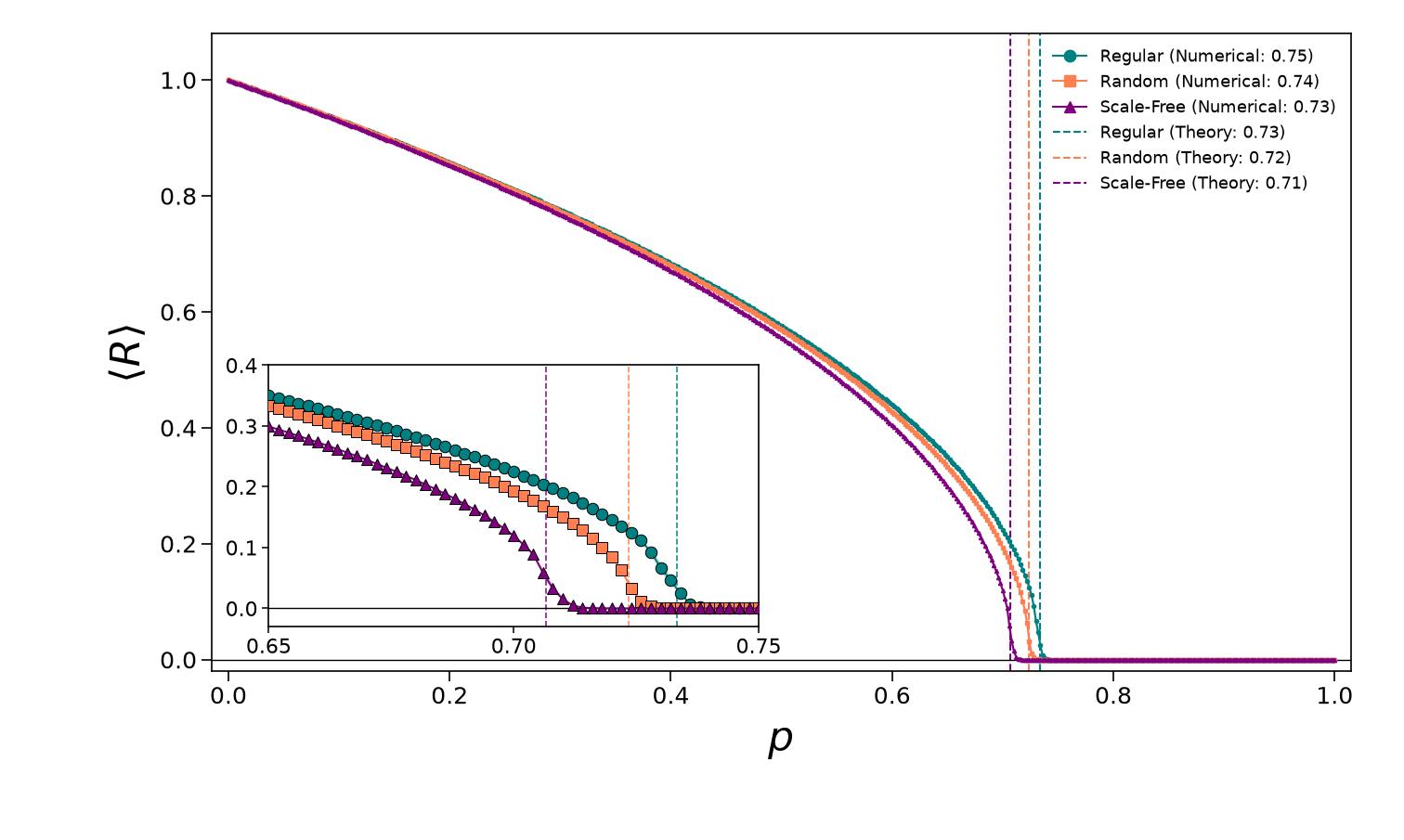}}
\caption{\textbf{Dynamical robustness of mobile Stuart-Landau agents across distinct network topologies.} The time-averaged macroscopic order parameter $\langle R \rangle$ is shown as a function of the fraction of unhealthy agents, $p$. Solid curves with markers denote the ensemble-averaged numerical results for agents moving on Regular (teal circles), Random Erdős-Rényi (coral squares), and Scale-Free Barabási-Albert (purple triangles) networks. Vertical dashed lines indicate the corresponding analytical critical thresholds, $p_c^{\mathrm{theory}}$, derived from the mean-field approximation. The numerical critical points, $p_c^{\mathrm{num}}$ (reported in the legend), mark the empirical onset of the amplitude death ($\langle R \rangle < 10^{-6}$). The results confirm that increased topological heterogeneity amplifies the effective global coupling, systematically shifting the critical transition point to lower values of $p$ and reducing the overall dynamical robustness of the system. Here, $N=500$, $n=50$, and $T_{\mathrm{total}} = 200$.}
\label{fig:aging_transition}
\label{Picture2}
\end{figure}

The set of mobile agents interacting with a given agent changes continuously because the agents perform random walks on the network.  When the mobility is sufficiently rapid compared with the evolution of the oscillator states, the instantaneous interaction network can be replaced by an effective, averaged global interaction. In the rapid-mobility regime, the interaction term in Eq.~\eqref{eq:microscopic_model} can be replaced by its temporal average. For an unbiased random walk on an undirected network, the stationary probability of finding an agent at node $\alpha$ is degree-weighted,
\begin{equation*}
P_\alpha=\frac{d_\alpha}{\sum_{\beta=1}^{n}d_\beta}
=\frac{d_\alpha}{n\langle d\rangle}.
\end{equation*}
%
% Hence, the probability that two independently moving agents occupy
% the same node is
% %
% \begin{equation}
% P_{\mathrm{same}}
% =\sum_{\alpha=1}^{n}P_\alpha^2
% =\frac{\langle d^2\rangle}
% {n\langle d\rangle^2}.
% \end{equation}
% %

For two independently moving agents, the long-time average of the co-location indicator is equal to the probability of occupying the same node,
\begin{equation}
\left\langle
\mathbf{1}_{\{x_i=x_j\}}
\right\rangle_t
=
P_{\mathrm{same}}
=
\sum_{\alpha=1}^{n}P_\alpha^2
=
\frac{\langle d^2\rangle}
{n\langle d\rangle^2}.
\label{eq:co_location_probability}
\end{equation}
%

% In the rapid-mobility regime, this temporal average can be used to
% replace the instantaneous co-location indicator in the coupling term,
% yielding
% %
% \begin{equation}
% k\sum_{j\neq i}
% \mathbf{1}_{\{X_i=X_j\}}(Z_j-Z_i)
% \simeq
% kP_{\mathrm{same}}
% \sum_{j\neq i}(Z_j-Z_i).
% \end{equation}
% %
% Since
% %
% \begin{equation}
% \sum_{j\neq i}(Z_j-Z_i)
% =
% N(\bar Z-Z_i),
% \end{equation}
% %
% the effective coupling is
% %
% \begin{equation}
% \widetilde{k}
% =
% NP_{\mathrm{same}}k
% =
% \frac{N}{n}
% \frac{\langle d^2\rangle}{\langle d\rangle^2}k.
% \end{equation}

%Thus, on average, an agent interacts with approximately
%$N P_{\mathrm{same}}$ other agents.

Replacing the instantaneous co-location indicator by this stationary average gives
\begin{equation*}
k\sum_{j\neq i}
\mathbf{1}_{\{x_i=x_j\}}(Z_j-Z_i)
\simeq
kNP_{\mathrm{same}}(\bar Z-Z_i),
\end{equation*}
where
$\bar Z=N^{-1}\sum_j Z_j$ is the population mean field. Therefore, the microscopic mobile system reduces, in the rapid-mobility limit, to a globally coupled Stuart--Landau population,
\begin{equation}
\frac{dZ_i}{dt}
=
(a_i+i\omega-|Z_i|^2)Z_i
+
\widetilde{k}(\bar Z-Z_i),
\label{main}
\end{equation}

\begin{figure*}[htp]
\centerline{\includegraphics[width=1.0\textwidth]{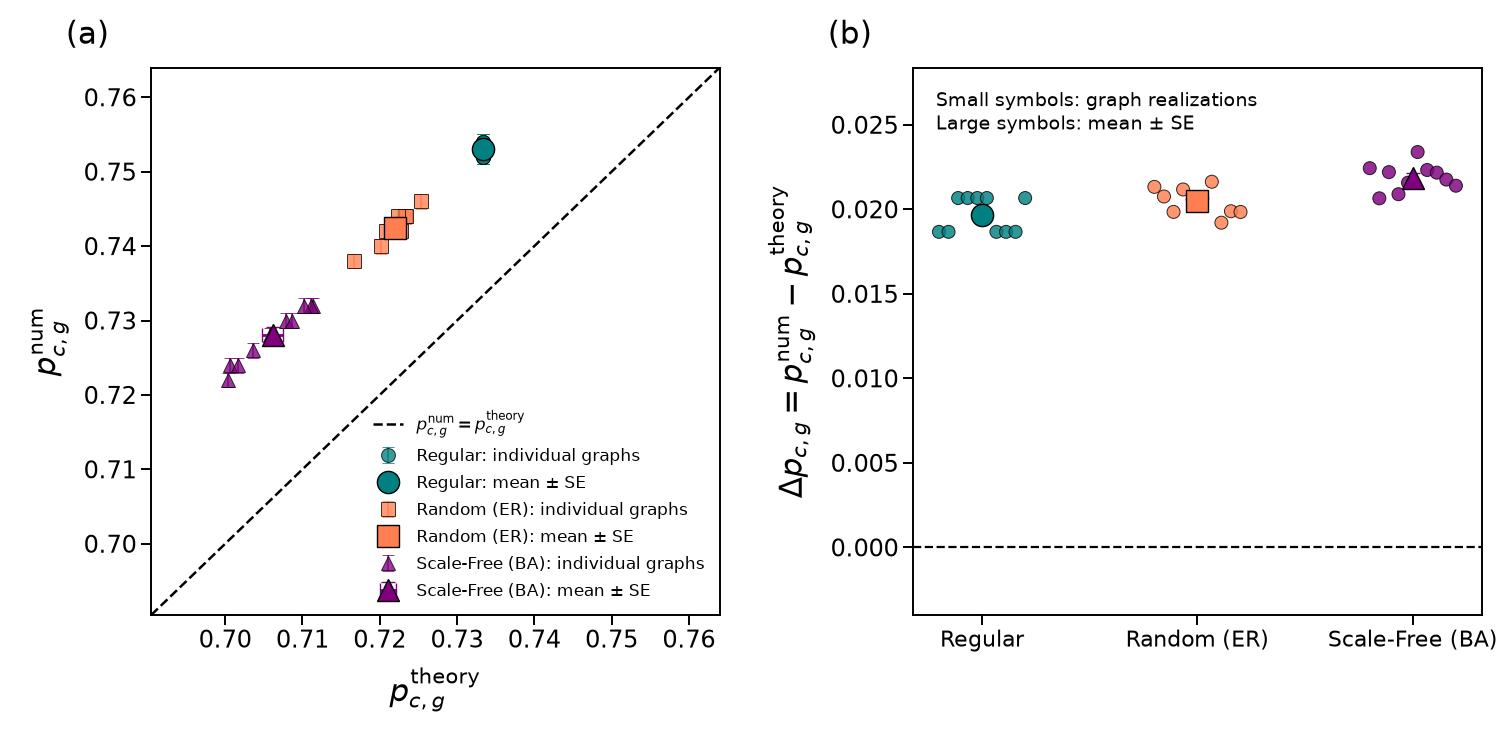}}
\caption{\textbf{Network-ensemble comparison of numerical and analytical aging thresholds.} (a) The numerically obtained critical point, $p_{c,g}^{\mathrm{num}}$, plotted against the mean-field theoretical prediction, $p_{c,g}^{\mathrm{theory}}$, for individual network realizations $g$ of Regular (teal circles), Random Erdős-Rényi (coral squares), and Scale-Free Barabási-Albert (purple triangles) topologies. The dashed black line indicates perfect theoretical agreement ($p_{c,g}^{\mathrm{num}} = p_{c,g}^{\mathrm{theory}}$). (b) The signed deviation, $\Delta p_{c,g} = p_{c,g}^{\mathrm{num}} - p_{c,g}^{\mathrm{theory}}$, exhibits a systematic positive shift of approximately $+0.02$ across all topologies. Despite this offset, the theoretical topology-dependent ordering, $p_c^{\mathrm{BA}} < p_c^{\mathrm{ER}} < p_c^{\mathrm{Regular}}$, is strictly preserved. In both panels, small symbols represent individual graph realizations ($10$ per topological class). For each fixed network and fraction $p$, the order parameter is evaluated by averaging over $10$ independent stochastic mobile-agent trajectories with different initial conditions and random-walk histories. Large symbols denote the ensemble mean over the $10$ network realizations. The numerical threshold is defined precisely as the first sampled $p$ for which $\langle R \rangle < 10^{-6}$. Because the parameter grid spacing is $\Delta p = 0.002$ and the graph-to-graph variance is relatively low, the standard error bars are visually entirely contained within the large mean markers. Here, $N=500$, $n=50$, and $T_{\mathrm{total}} = 200$.}
\label{Picture4}
\end{figure*}

% \begin{figure*}[htp]
% \centerline{\includegraphics[width=1.0\textwidth]{Figure_4.jpeg}}
% \caption{\textbf{Network-ensemble comparison of numerical and analytical aging thresholds.} (a) The numerically obtained critical point, $p_{c,g}^{\mathrm{num}}$, plotted against the mean-field theoretical prediction, $p_{c,g}^{\mathrm{theory}}$, for Regular (teal circles), Random Erdős--Rényi (coral squares), and Scale-Free Barabási--Albert (purple triangles) networks. The dashed black line indicates perfect theoretical agreement ($p_{c,g}^{\mathrm{num}} = p_{c,g}^{\mathrm{theory}}$). (b) The signed deviation, $\Delta p_{c,g} = p_{c,g}^{\mathrm{num}} - p_{c,g}^{\mathrm{theory}}$, exhibiting a systematic positive shift of approximately $+0.02$ across all topologies due to finite-size stochastic fluctuations. In both panels, small symbols represent individual, independent graph realizations ($10$ per topological class), while large symbols denote the ensemble mean. Error bars represent the standard error of the mean across the graph realizations; they are visually entirely contained within the large markers due to the exceptionally low graph-to-graph variance and the discrete parameter grid resolution ($\Delta p = 0.002$).}
% \label{Picture4}
% \end{figure*}

where the resulting effective coupling strength \cite{mikaberidze2024consensus} is
\begin{equation}
\boxed{
\widetilde{k}
= kNP_{\mathrm{same}} 
= \frac{N}{n}
\frac{\langle d^2\rangle}{\langle d\rangle^2}
k
}
\label{eq:effective_coupling}
\end{equation}
for identical microscopic coupling strength $k$.
Here,
\begin{equation*}
\langle d\rangle
=
\frac{1}{n}
\sum_{\alpha=1}^{n} d_\alpha
\end{equation*}
is the mean degree of the network, and
\begin{equation*}
\langle d^2\rangle
=
\frac{1}{n}
\sum_{\alpha=1}^{n} d_\alpha^2
\end{equation*}
is the second moment of the degree distribution.  The ratio
\begin{equation*}
\frac{\langle d^2\rangle}{\langle d\rangle^2}
\end{equation*}
measures the enhancement of the effective coupling produced by degree heterogeneity. For a perfectly homogeneous network, where every node has the same
degree $d$, one has
\begin{equation*}
\langle d\rangle=d,
\qquad
\langle d^2\rangle=d^2,
\end{equation*}
and therefore
\begin{equation*}
\widetilde{k}
=
\frac{N}{n}k.
\end{equation*}

For a heterogeneous network, the degree variance is nonzero and
\begin{equation*}
\langle d^2\rangle
>
\langle d\rangle^2.
\end{equation*}
Thus, for fixed $N$, $n$, and $k$, degree heterogeneity increases $\widetilde{k}$. Equation~\eqref{eq:effective_coupling} is the key connection between
the underlying network topology and the effective globally coupled oscillator model.
%After averaging over the rapid mobility, Eq.~\eqref{eq:microscopic_model}
% takes the effective form
% %
% \begin{equation}
% \frac{dZ_i}{dt}
% =
% \left(a_i+i\omega-|Z_i|^2\right)Z_i
% +
% \frac{\widetilde{k}}{N}
% \sum_{j=1}^{N}
% \left(Z_j-Z_i\right).
% \label{eq:effective_model}
% \end{equation}

% Introducing the global complex mean field
% %
% \begin{equation}
% \overline{Z}
% =
% \frac{1}{N}
% \sum_{j=1}^{N}Z_j,
% \label{eq:global_mean}
% \end{equation}
%
% the coupling term can be written as
% %
% \begin{equation}
% \frac{\widetilde{k}}{N}
% \sum_{j=1}^{N}(Z_j-Z_i)
% =
% \widetilde{k}\left(\overline{Z}-Z_i\right).
% \end{equation}

% Therefore, 
% the effective model becomes
% %
% \begin{equation}
% % \boxed{
% \frac{dZ_i}{dt}
% =
% \left(a_i+i\omega-|Z_i|^2\right)Z_i
% +
% \widetilde{k}\left(\overline{Z}-Z_i\right).
% % }
% \label{eq:effective_meanfield}
% \end{equation}

% This equation has a simple interpretation.  Each oscillator evolves
% according to its own intrinsic Stuart-Landau dynamics while being
% attracted toward the instantaneous global mean field
% $\overline{Z}$.  The parameter $\widetilde{k}$ determines the strength
% of this global interaction.

% ------------------------------------------------------------
%\subsection{Two-population description}

\begin{figure*}[htp]
\centerline{\includegraphics[width=1.0\textwidth]{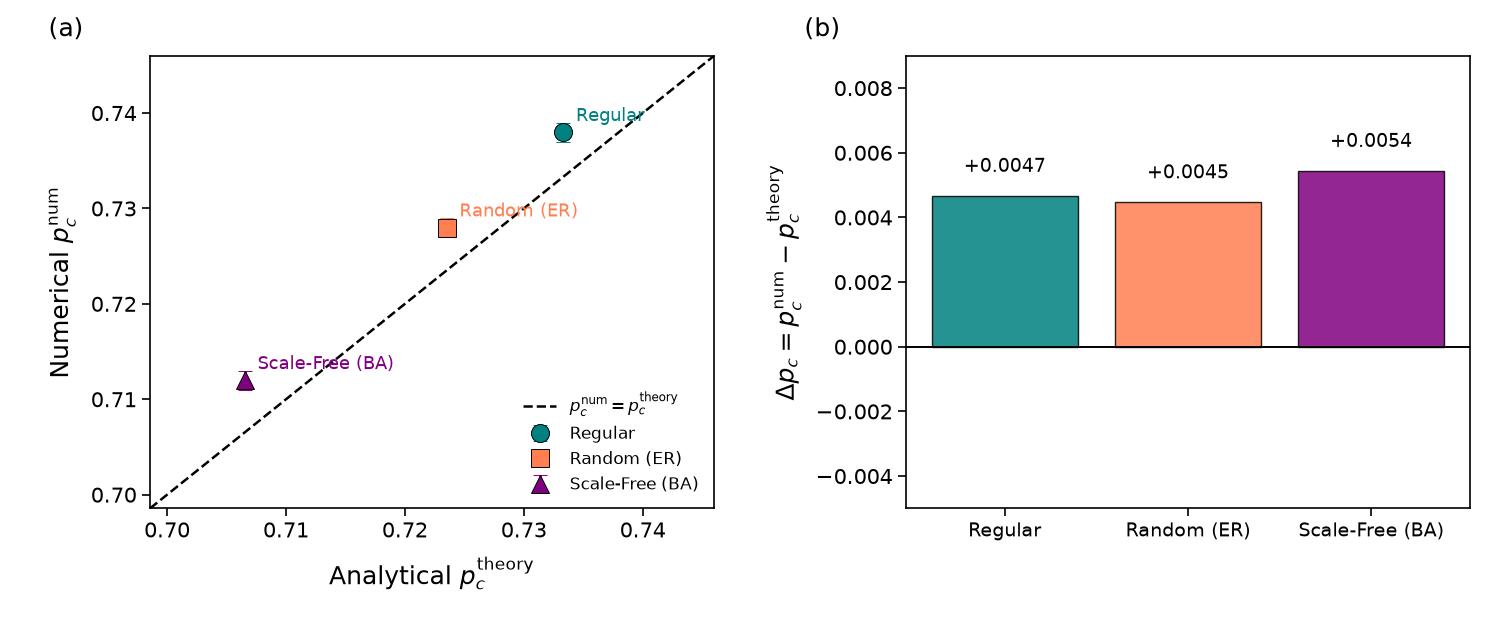}}
\caption{\textbf{Validation of the analytical critical thresholds.}
 (a) The numerically obtained critical transition point, $p_c^{\mathrm{num}}$, plotted against the mean-field analytical prediction, $p_c^{\mathrm{theory}}$, for mobile agents on Regular (teal circles), Random Erdős-Rényi (coral squares), and Scale-Free Barabási-Albert (purple triangles) networks. For each network topology, a single network realization is generated and kept fixed throughout the simulations. For each \(p\), the dynamics are averaged over $10$ independent stochastic mobile agents' trajectories. The dashed black line represents exact agreement $p_c^{\mathrm{num}} = p_c^{\mathrm{theory}}$. $p_c^{\rm num}$ is defined as the first sampled $p$ for which the ensemble-averaged order parameter satisfies $\langle R\rangle<10^{-6}$. (b) The difference, $\Delta p_c = p_c^{\mathrm{num}} - p_c^{\mathrm{theory}}$, across the three topological classes. 
 % {\color{red}Error bars indicate half the numerical grid spacing, $\Delta p/2=0.001$, and therefore represent the resolution with which the threshold can be localized on the discrete $p$ grid}.
 A small positive deviation, $\Delta p_c \simeq 0.005$, remains across the three topologies. Extending the total integration time from $T_{\mathrm{total}}=200$ to $T_{\mathrm{total}}=2000$ substantially reduces this discrepancy, indicating that finite-time relaxation effects contribute appreciably to the observed offset. Despite these slight positive deviation relative to the rapid-mobility averaged theory, the theoretical framework perfectly predicts the topological hierarchy of the systemic vulnerability, confirming that structural heterogeneity definitively lowers the critical threshold. Here, $N=500$, $n=50$, and $T_{\mathrm{total}} = 2000$.}
\label{Picture3}
\end{figure*}

% \begin{itemize}
%     \item a healthy population with intrinsic parameter
%     $a_{\mathrm{healthy}}>0$;
%     \item an unhealthy population with an intrinsic parameter
%     $a_{\mathrm{unhealthy}}<0$.
% \end{itemize}

We now divide the population into two groups: (i) a healthy population exhibiting stable oscillation with intrinsic parameter $a_{\mathrm{healthy}}>0$; and (ii) an unhealthy population resting at the quiescent state with an intrinsic parameter $a_{\mathrm{unhealthy}}<0$. Let $p \in [0,1]$ denote the fraction of unhealthy agents.  Therefore,
\begin{equation*}
p
=
\frac{N_{\mathrm{unhealthy}}}{N},
\end{equation*}

and the fraction of healthy agents is
\begin{equation*}
1-p
=
\frac{N_{\mathrm{healthy}}}{N}.
\end{equation*}

Let $Z_{healthy}$ denote the complex mean field of the healthy group and $Z_{unhealthy}$ denote the complex mean field of the unhealthy group. The total mean field is then
\begin{equation}
% \boxed{
\bar{Z}
=
(1-p)Z_{healthy}+pZ_{unhealthy}.
% }
\label{eq:population_mean}
\end{equation}

Substituting Eq.~\eqref{eq:population_mean} into the effective Stuart-Landau equation \eqref{main} gives the equations for the two populations.

For the healthy population,
% \begin{align}
% \frac{dZ_A}{dt}
% % &=
% % \left(a_{\mathrm{active}}+i\omega\right)Z_A
% % +
% % \widetilde{k}
% % \left[
% % (1-p)Z_A+pZ_G-Z_A
% % \right]
% % -
% % |Z_A|^2Z_A
% % \nonumber\\
% &=
% \left(
% a_{\mathrm{healthy}}
% -\widetilde{k}p
% +i\omega
% \right)Z_{healthy}
% +
% \widetilde{k}pZ_{unhealthy}
% -
% |Z_{healthy}|^2Z_{healthy}.
% \label{eq:active_population}
% \end{align}
\begin{align}
\frac{dZ_{\mathrm{healthy}}}{dt} &= \left(a_{\mathrm{healthy}} - \widetilde{k}p + i\omega\right)Z_{\mathrm{healthy}} \nonumber \\
&\quad + \widetilde{k}p Z_{\mathrm{unhealthy}} - |Z_{\mathrm{healthy}}|^2 Z_{\mathrm{healthy}}.
\label{eq:active_population}
\end{align}

Similarly, for the unhealthy population,
% \begin{align}
% \frac{dZ_{unhealthy}}{dt}
% % &=
% % \left(a_{\mathrm{aged}}+i\omega\right)Z_G
% % +
% % \widetilde{k}
% % \left[
% % (1-p)Z_A+pZ_G-Z_G
% % \right]
% % -
% % |Z_G|^2Z_G
% % \nonumber\\
% &=
% \widetilde{k}(1-p)Z_{healthy}
% +
% \left(
% a_{\mathrm{aged}}
% -\widetilde{k}(1-p)
% +i\omega
% \right)Z_{unhealthy}
% -
% |Z_{unhealthy}|^2Z_{unhealthy}.
% \label{eq:aged_population}
% \end{align}
\begin{align}
\frac{dZ_{\mathrm{unhealthy}}}{dt} &= \widetilde{k}(1-p)Z_{\mathrm{healthy}} \nonumber \\
&\quad + \left(a_{\mathrm{unhealthy}} - \widetilde{k}(1-p) + i\omega\right)Z_{\mathrm{unhealthy}} \nonumber \\
&\quad - |Z_{\mathrm{unhealthy}}|^2 Z_{\mathrm{unhealthy}}.
\label{eq:aged_population}
\end{align}

% Equations~\eqref{eq:active_population} and
% \eqref{eq:aged_population} describe the competition between the
% self-sustained active population and the quiescent aged population.

% ------------------------------------------------------------
% \subsection{Linear stability of the quiescent state}

% The critical aged fraction $p_c$ is determined by the loss of
% stability of the macroscopic oscillatory state.  Equivalently, we
% can examine the stability of the quiescent state

Examining the linear stability of the quiescent state
\begin{equation*}
Z_{healthy}=Z_{unhealthy}=0,
\end{equation*}

we obtain the analytical value of the critical fraction of unhealthy agents \cite{daido2004aging}
\begin{equation}
\boxed{
p_c
=
\frac{
a_{\mathrm{healthy}}
\left(
\widetilde{k}-a_{\mathrm{unhealthy}}
\right)
}{
\widetilde{k}
\left(
a_{\mathrm{healthy}}-a_{\mathrm{unhealthy}}
\right)
}.
}
\label{eq:pc_general}
\end{equation}

This expression gives the maximum fraction of unhealthy agents that can be present while the macroscopic oscillatory state remains dynamically sustainable under the effective globally coupled model.

To analytically determine how network heterogeneity influences dynamical robustness, we differentiate the critical threshold $p_c$ with respect to the effective global coupling $\widetilde{k}$:
\begin{equation*}
\frac{d p_c}{d \widetilde{k}}
=
\frac{a_{\mathrm{healthy}} a_{\mathrm{unhealthy}}}{\widetilde{k}^2 \left( a_{\mathrm{healthy}} - a_{\mathrm{unhealthy}} \right)}.
\label{eq:pc_derivative}
\end{equation*}
By definition, $a_{\mathrm{healthy}} > 0$ and $a_{\mathrm{unhealthy}} < 0$ are parameters. Therefore, the numerator is strictly negative while the denominator is strictly positive, yielding
\begin{equation*}
\frac{d p_c}{d \widetilde{k}} < 0.
\label{eq:pc_inequality}
\end{equation*}
This proves that any increase in $\widetilde{k}$, such as the amplification caused by topological hubs in scale-free networks, directly reduces the critical threshold. Consequently, stronger mobility-mediated coupling makes the collective oscillatory state inherently more vulnerable to functional degradation.

This reduction in robustness originates from the diffusive nature of the coupling. Diffusive coupling does not generate oscillatory activity; it reduces differences between interacting oscillator states. When healthy oscillators interact with other healthy oscillators that are already synchronized, $Z_j - Z_i \simeq 0$, so the coupling contribution is negligible, and the oscillators retain their intrinsic amplitude $\sqrt{a_{\mathrm{healthy}}}$. In contrast, when a healthy oscillator encounters an unhealthy one, whose state is driven toward $Z=0$, the coupling term becomes nonzero and pulls the healthy oscillator toward the quiescent state. The reverse influence is also present: healthy oscillators drive unhealthy ones away from the origin. However, because unhealthy oscillators have $a_{\mathrm{unhealthy}} < 0$, their intrinsic dynamics continuously restore them towards $Z=0$; consequently, increasing the coupling does not simply transfer activity from healthy to unhealthy population, but increasingly enforces a common state between active and intrinsically damped oscillators. Network heterogeneity amplifies this effect because random walkers preferentially visit high-degree nodes, increasing the probability of co-location and hence the effective coupling. Thus, hubs do not preferentially transmit inactivity; rather, they increase the rate at which the two populations are coupled. Since healthy-healthy interactions largely vanish once synchrony is established, whereas healthy-unhealthy interactions remain dynamically mismatched, the increased coupling disproportionately matters for suppressing the collective oscillation. This physically explains why a larger effective coupling lowers the critical fraction of unhealthy agents required for amplitude death \cite{ad_report,koseska2013oscillation,dixit2021emergent,karnatak2007amplitude,mirollo1990amplitude,dixit2021dynamic,nonlinear}.

% ------------------------------------------------------------
% \subsection{Analytical expression for the parameters used here}

% To quantify the collective oscillation, we define the instantaneous
% complex mean field as

% \begin{equation}
% \overline{Z}(t)
% =
% \frac{1}{N}
% \sum_{i=1}^{N} Z_i(t).
% \end{equation}

To quantify the collective oscillation, we define the instantaneous normalized order parameter as
\begin{equation*}
R(t)
=
\frac{|\bar{Z}(t)|}
{\sqrt{a_{\mathrm{healthy}}}}.
\label{eq:instantaneous_order_parameter}
\end{equation*}

% The normalization by $\sqrt{a_{\mathrm{active}}}$ is convenient
% because an isolated active Stuart--Landau oscillator has amplitude
% $\sqrt{a_{\mathrm{active}}}$.
When all agents are healthy and sufficiently synchronized, $R(t)$ is of order unity.

Numerical simulations are performed using a population of $N=500$ agents moving across networks of $n=50$ nodes with a target mean degree of $\langle d \rangle = 4$. The spatial environments are generated as regular, Erdős-Rényi (ER) random, and Barabási-Albert (BA) scale-free networks. The regular network has exactly $\langle d\rangle=4$, while the ER and BA networks exhibit realization-dependent mean degrees and degree heterogeneity. For each network topology, a single network realization is generated and kept fixed throughout the simulations (unless stated otherwise). For each value of $p$, the dynamics are averaged over $10$ independent stochastic realizations of the mobile-agent trajectories on the same fixed network, with each realization differing in its initial conditions and random-walk history. Error bars denote the standard error across these realizations. Agent locations $x_i(0)$ are initially assigned uniformly at random across all nodes, and initial complex states $Z_i(0)$ are drawn from a uniform distribution bounded by $[-1, 1]$ in both the real and imaginary axes. The microscopic continuous-time dynamics are integrated using a fourth-order Runge-Kutta method (RK4) with a fixed integration time step of $dt=0.01$. To capture the rapid-mobility regime, agents update their spatial locations via a random walk, moving to a uniformly selected neighboring node, at discrete time intervals of $\Delta T = 0.01$. Each simulation is evolved for a total time of $T_{\mathrm{total}} = 200$. To eliminate transient effects, the collective macroscopic order parameter is calculated as a time average over the final $T_{\mathrm{average}}=50$ time units of the integration. Therefore, the time-averaged order parameter is given by

\begin{equation}
%\boxed{
\langle {R} \rangle
=
\frac{1}{T_{\mathrm{average}}}
\int_{T_{\mathrm{total}}-T_{\mathrm{average}}}^{T_{\mathrm{total}}}
R(t)\,dt.
%}
\label{eq:time_average_order_parameter}
\end{equation}

The system parameters are fixed at intrinsic frequency $\omega = 3.0$, coupling strength $k = 1.0$, and parameters $a_{\mathrm{healthy}} = 2.0$ and $a_{\mathrm{unhealthy}} = -1.0$.

% Rather than measuring the order parameter at a single final time,
% we use the time average over the final $T_{\mathrm{average}}$
% units of the simulation:

% \begin{equation}
% \boxed{
% \overline{R}_{T_{\mathrm{average}}}
% =
% \frac{1}{T_{\mathrm{average}}}
% \int_{T_{\mathrm{total}}-T_{\mathrm{average}}}^{T_{\mathrm{total}}}
% R(t)\,dt.
% }
% \label{eq:time_average_order_parameter}
% \end{equation}

% We use adaptive RK45 integration with maximum step size $dt=0.01$, while agents performed random walks every $\Delta T=0.01$. Simulations are done for $T_{total}=200$, with the order parameter averaged over the final $50$ time units and over $10$ independent realizations for each $p$. 

%To account for finite-size effects and stochastic fluctuations, the reported order parameter at each aged fraction $p$ represents the ensemble mean over $10$ independent random realizations (varying initial agent states, positions, and random walk trajectories), with the corresponding standard error used to quantify statistical uncertainty.

% In our simulations, we choose

% \begin{equation}
% a_{\mathrm{healthy}}=2,
% \qquad
% a_{\mathrm{unhealthy}}=-1.
% \end{equation}

Substituting these values into Eq.~\eqref{eq:pc_general} gives

% \begin{align}
% p_c
% &=
% \frac{
% 2(\widetilde{k}-(-1))
% }{
% \widetilde{k}(2-(-1))
% }
% \\
% &=
% \frac{2(\widetilde{k}+1)}
% {3\widetilde{k}}.
% \end{align}

% Therefore,

\begin{equation}
%  \boxed{
p_c
=
\frac{2}{3}
\left(
1+\frac{1}{\widetilde{k}}
\right)
=
\frac{2}{3}
\left[
1+
\frac{1}{
\displaystyle
\frac{N}{n}
\frac{\langle d^2\rangle}
{\langle d\rangle^2}
k
}
\right].
% }
\label{eq:pc_specific}
\end{equation}

% Using the effective coupling from Eq.~\eqref{eq:effective_coupling},

% \begin{equation}
% \widetilde{k}
% =
% \frac{N}{n}
% \frac{\langle d^2\rangle}
% {\langle d\rangle^2}
% k,
% \end{equation}

% Using the effective coupling from Eq.~\eqref{eq:effective_coupling}, the critical fraction can also be written entirely in terms of the
% network statistics and the microscopic parameters:

% \begin{equation}
% % \boxed{
% p_c
% =
% \frac{2}{3}
% \left[
% 1+
% \frac{1}{
% \displaystyle
% \frac{N}{n}
% \frac{\langle d^2\rangle}
% {\langle d\rangle^2}
% k
% }
% \right].
% % }
% \label{eq:pc_network}
% \end{equation}

Thus, the network topology enters the analytical prediction through the degree moments $\langle d\rangle$ and $\langle d^2\rangle$.

% ------------------------------------------------------------
%\subsection{Example for the present simulation}

% For the simulations considered here,

% \begin{equation}
% N=100,
% \qquad
% n=20,
% \qquad
% k=1.
% \end{equation}

% Therefore,

% \begin{equation}
% \frac{N}{n}=5,
% \end{equation}

% and

% \begin{equation}
% \widetilde{k}
% =
% 5
% \frac{\langle d^2\rangle}
% {\langle d\rangle^2}.
% \end{equation}

For a regular network with degree $d=4$, we have $\langle d\rangle=4,$ and $
\langle d^2\rangle=16,$ so that $\widetilde{k}_{\mathrm{Regular}}
=
10.$ Consequently,

% \begin{equation}
% \langle d\rangle=4,
% \qquad
% \langle d^2\rangle=16,
% \end{equation}

% so that

% \begin{equation}
% \widetilde{k}_{\mathrm{Regular}}
% =
% 10\frac{16}{16}
% =
% 10.
% \end{equation}

% Consequently,

\begin{equation*}
p_c^{\mathrm{Regular}}
=
\frac{2}{3}
\left(
1+\frac{1}{10}
\right)
\approx
0.7333.
\end{equation*}

In contrast, Erd\H{o}s-R\'enyi networks follow a Poisson degree distribution, whereas Barab\'asi-Albert networks exhibit power-law degree distribution with pronounced degree heterogeneity. In both cases, however the degree variance is nonzero, giving $\langle d^2\rangle
>
\langle d\rangle^2.$ Consequently,

% \begin{equation}
% \langle d^2\rangle
% >
% \langle d\rangle^2.
% \end{equation}

% Consequently,

\begin{equation*}
\widetilde{k}_{\mathrm{heterogeneous}}
>
\frac{N}{n}k = \widetilde{k}_{\mathrm{Regular}},
\end{equation*}

which leads, through Eq.~\eqref{eq:pc_specific}, to a smaller analytical critical fraction than $p_c^{\mathrm{Regular}}$. The dependence of $p_c$ on the effective coupling can be obtained
directly from Eq.~\eqref{eq:pc_specific}.

%%%%%%%%%%%%%%%%%%%%%%%%%%%%%%%%%%%%%%%%%%%%%%%%%%%%%%%%%%%%%%%

Figure~\ref{fig:aging_transition} shows the dynamical robustness for mobile oscillators on Regular, Erdős-Rényi (ER), and Barabási-Albert (BA) mobility networks. In all three cases, the collective order parameter $\langle R\rangle$ decreases continuously with increasing fraction of unhealthy agents and eventually collapses to zero. The transition shifts systematically toward lower $p$ as network heterogeneity increases, with the scale-free network losing collective oscillations first, followed by the ER and Regular networks. The analytical thresholds (dashed lines) closely capture this topology-dependent shift, demonstrating that the enhanced effective coupling induced by degree heterogeneity reduces the tolerance of the collective state to functional degradation. The numerical thresholds exhibit a slight systematic positive deviation ($\Delta p_c \approx +0.02$) relative to the analytical predictions of the mean-field of rapid-mobility in all three topologies.

To assess the robustness of the analytical threshold against finite-network size and stochastic mobility fluctuations, we further perform an ensemble calculation using $10$ independent network realizations for each topology. Each network realization is generated independently and then kept fixed throughout the corresponding simulations. To account for stochasticity in the agent mobility, the time-averaged order parameter for each fixed network and fraction $p$ is further ensemble-averaged over $10$ independent random-walk trajectories with unique initial oscillator states and initial mobile agents' locations. %The microscopic continuous-time dynamics are integrated using a fixed-step fourth-order Runge-Kutta (RK4) numerical scheme with integration step $ dt=0.01$. 
The order parameter \eqref{eq:time_average_order_parameter} is averaged over the final $T_{\mathrm{average}}=50$ time units of a total simulation time $T_{\mathrm{total}}=200$. Since the fraction of unhealthy agents is varied in increments of $1/N=0.002$, the numerical transition point is then strictly defined as the first parameter step on a discrete grid ($\Delta p = 0.002$) where the macroscopic order parameter drops below a stringent amplitude-death criterion of $\langle R \rangle < 10^{-6}$.

% For each network realization $g$, the numerical critical fraction $p_{c,g}^{\mathrm{num}}$ was defined as the first sampled value of $p=n_{\mathrm{aged}}/N$ for which the mobility-ensemble-averaged order parameter satisfies
% \begin{equation}
% \langle R\rangle_{\mathrm{walk}} < 10^{-6}.
% \end{equation}
% Since the fraction of unhealthy agents is varied in increments of $1/N=0.002$, the corresponding threshold has a grid resolution of $\Delta p/2=0.001$. The analytical threshold was evaluated independently for each network realization from its realized degree moments,
% \begin{equation}
% \tilde{k}_{g}
% =
% \frac{N}{n}
% \frac{\langle d^2\rangle_{g}}
% {\langle d\rangle_{g}^{2}}\,k,
% \end{equation}
% giving
% \begin{equation}
% p_{c,g}^{\mathrm{theory}}
% =
% \frac{
% a_{\mathrm{healthy}}
% \left(\tilde{k}_{g}-a_{\mathrm{unhealthy}}\right)
% }{
% \tilde{k}_{g}
% \left(a_{\mathrm{healthy}}-a_{\mathrm{unhealthy}}\right)
% }.
% \end{equation}

For each network realization $g$, we quantify the network-level discrepancy between the numerical and analytical thresholds by
\begin{equation*}
\Delta p_{c,g}
=
p_{c,g}^{\mathrm{num}}
-
p_{c,g}^{\mathrm{theory}}.
\end{equation*}

% For an observable $x_g$ evaluated over the $N_G=10$ independent network realizations, the ensemble mean and standard error (SE) are calculated as
% \begin{equation}
% \bar{x}
% =
% \frac{1}{N_G}
% \sum_{g=1}^{N_G}x_g,
% \end{equation}
% and
% \begin{equation}
% \mathrm{SE}(x)
% =
% \frac{s_x}{\sqrt{N_G}},
% \qquad
% s_x
% =
% \sqrt{
% \frac{1}{N_G-1}
% \sum_{g=1}^{N_G}
% \left(x_g-\bar{x}\right)^2
% }.
% \end{equation}
% Thus, the error bars associated with the topology-level values of $\Delta p_c$ quantify the uncertainty in the mean across network realizations,
% \begin{equation}
% \mathrm{SE}(\Delta p_c)
% =
% \frac{s_{\Delta p_c}}{\sqrt{N_G}}.
% \end{equation}

Figure~\ref{Picture4} summarizes this network-ensemble validation and shows that the numerical thresholds systematically lie above the rapid-mobility predictions, while the topology-dependent ordering
\begin{equation*}
p_c^{\mathrm{BA}}
<
p_c^{\mathrm{ER}}
<
p_c^{\mathrm{Regular}}
\end{equation*}
is consistently preserved across network realizations. This demonstrates that the analytical theory captures the topology-dependent shift of the transition point, while the remaining systematic offset at $T_{\rm total}=200$ can be attributed to finite-time, finite-population, and finite-mobility effects not explicitly captured by the rapid-mobility mean-field approximation. The relatively small standard errors reflect the narrow network-to-network dispersion of $\Delta p_{c,g}$ across the $10$ network realizations. The standard error of the mean
discrepancy is calculated as
\[
SE(\Delta p_c)
=
\frac{s_{\Delta p_c}}{\sqrt{N_g}},
\qquad N_g=10,
\]
where $s_{\Delta p_c}$ is the sample standard deviation of $\Delta p_{c,g}$ across the $N_g$ network realizations.
%The small standard errors therefore reflect the relatively narrow graph-to-graph dispersion of the threshold discrepancy.

%; they quantify the uncertainty of the ensemble mean rather than the stochastic trajectory-to-trajectory variability within an individual network.

% As shown in Fig.~\ref{fig:network_ensemble}, the numerically evaluated thresholds exhibit a systematic, topology-independent positive deviation ($\Delta p_{c,g} \approx +0.02$) compared to the infinite-size mean-field theory[cite: 4]. This marginal delay in complete amplitude death is a standard hallmark of finite-size stochastic fluctuations inherent to the discrete $N=500$ agent population. Crucially, despite this microscopic offset, the analytical framework perfectly predicts the topological hierarchy of vulnerability: $p_c^{\mathrm{BA}} < p_c^{\mathrm{ER}} < p_c^{\mathrm{Regular}}$[cite: 4]. To quantify structural variance, we compute the standard error of the mean across the $10$ independent graph realizations for each topology[cite: 4]. Because the critical transition point is highly deterministic across randomly generated networks of the same class, and due to the rigid spacing of the discrete parameter grid, the graph-to-graph variance is exceptionally low[cite: 4]. Consequently, the resulting standard errors are microscopically small (on the order of $10^{-4}$) and are visually entirely subsumed within the ensemble-mean markers, confirming the high robustness and reproducibility of the topological vulnerability mechanism[cite: 4].

To examine the possible reasons behind this marginal discrepancy in between theoretical and numerical results, we increase the total integration time to $T_{\mathrm{total}}=2000$ and compute the collective order parameter by averaging over the final $T_{\mathrm{average}}=500$ time units. As shown in Fig.~\ref{Picture3}, the resulting thresholds remain close to the analytical predictions, with deviations of only $0.0045$-$0.0054$. The reduction of the discrepancy from $\Delta p_c \simeq 0.02$ at $T = 200$ to $\Delta p_c \simeq 0.005$ at $T = 2000$ indicates that finite simulation time accounts for a substantial fraction of the initial offset. This close agreement confirms the robustness of the analytical prediction, while the small residual offset is attributable to the finite-resolution, finite-time, finite-population, and stochastic nature of the simulations not captured by the rapid-mobility mean-field
approximation.

We have demonstrated that mobility fundamentally reshapes the dynamical robustness of oscillator populations undergoing functional degradation. By mapping the rapidly fluctuating co-location dynamics onto an effective mean-field description, we show analytically that network degree heterogeneity amplifies the effective global coupling. Because the critical fraction decreases monotonically with this effective coupling, degree heterogeneity lowers the analytical threshold for the aging transition. %By mapping discrete random walks into an effective mean-field description, we analytically prove that network degree variance amplifies the global interaction strength among mobile Stuart-Landau oscillators. Because the derivative of the critical threshold with respect to this effective coupling is unconditionally negative, this topological amplification strictly lowers the threshold for the aging transition.
Consequently, mobile swarms traversing scale-free networks suffer collective amplitude death significantly earlier than those on homogeneous regular or random topologies. High-precision numerical simulations rigorously validate the theoretically predicted ordering, $p_c^{\mathrm{BA}} < p_c^{\mathrm{ER}} < p_c^{\mathrm{Regular}}$, confirming that structural heterogeneity degrades dynamical robustness in this mobility-mediated aging mechanism. Broadly, these results establish network-mediated mobility as a critical control parameter for the resilience of collective oscillations.

Building on this framework, several natural avenues for future research emerge. A crucial next step is to relax the rapid-mobility assumption and investigate intermediate hopping rates comparable to intrinsic oscillator time scales. This would provide a natural extension toward a quantitative theory of the transition outside the rapid-mobility regime. Furthermore, investigating biased navigation, where agents preferentially migrate toward active neighbors or avoid functionally degraded unhealthy neighbors, could reveal whether active matter can self-organize to mitigate these topological vulnerabilities. The mean-filed reduction in the rapid-mobility regime provides a general framework for incorporating network structure into the effective dynamics of mobile oscillators. Whether the resulting relationship between network heterogeneity and robustness threshold persists across other classes of nonlinear oscillators remains an open question, which can be tested by extending our analytical mapping beyond the Hopf bifurcation to chaotic systems like R\"ossler oscillators \cite{rossler1976equation}. Finally, translating these insights from synthetic graphs to empirical spatial topologies, such as urban transportation grids, drone communication arrays, or biological swarms, offers a promising path toward developing predictive tools to prevent cascading functional failures in real-world physical and engineering systems.

\typeout{}
\bibliographystyle{apsrev4-1}
\bibliography{biblo}

\end{document}